# Study of the effect of semi-infinite crystalline electrodes on transmission of gold atomic wires using DFT

Abdul Sattar*[a], Raja Junaid Amjad*[a], Sumaira Yasmeen[a], Hafsa Javed[a], Hamid Latif, Hasan Mahmood[a], Azmat Iqbal[a], Arslan Usman[a], Majid Niaz Akhtar[a], Salman Naeem Khan[a], M R Dousti[b]

[a] COMSATS Institute of Information Technology Lahore 54000 Pakistan
[b] Laboratório de Espectroscopia de Materiais Funcionais (LEMAF), Instituto de Fisica de Sao Carlos, Universidade de SaoPaulo, Av. Trabalhador Saocarlense Sao Carlos-SP, Brazil

*Corresponding authors: sattar82@gmail.com
rajajunaid25@gmail.com

First principle calculations of the conductance of gold wires containing 3-8 atoms each with 2.39 Å bond length were performed using density functional theory. Three different configuration of wire/electrodes were used.  For zigzag wire with semi-infinite crystalline electrodes, even-odd oscillation is observed which is consistent with the previously reported results. A lower conductance was observed for the chain in semi-infinite crystalline electrodes compared to the chains suspended in wire-like electrode.  The calculated transmission spectrum for the straight and zig-zag wires suspended between semi-infinite crystalline electrodes showed suppression of transmission channels due to electron scattering occurring at the electrode-wire interface.

## Introduction

Atomic wires have fundamental interests in low-dimensional physics and technological applications such as molecular electronic devices [1, 2]. Electron transports through metallic monoatomic nanowires have attracted much attention because they are ultimately the basic building block for nanoelectronic devices [3]. Metallic nanowires can be formed by gently breaking the metallic contact by using remarkably simple experimental techniques like scanning tunneling microscope (STM) and mechanically controllable break junction (MCBJ) techniques [4]. Atomic wires have been considered as a crucial source of conductance in nano-electronics by last few decades [5]. Due to their specific structural and conduction properties, monoatomic gold nanowires (AuNWs) have attracted a considerable interest both by theories [6, 7] and experiments [8, 9, 10]. AuNWs are considered to be the ideal candidates as interconnects for linking molecular devises because of their 1D geometry and superior conducting properties [11, 12, 13]. Moreover, AuNWs have also promising applications in many areas like photonic [14], plasmonic [15] and sensing devices [16, 17, 18, 19, 20]. Molecular-dynamics calculations [21], predicted the formation of free-standing string of gold atoms between two bulk electrodes. It was also confirmed by experiments carried out by high-resolution transmission electron microscope [22], and MCBJ [23]. The even-odd numbered oscillatory behavior of conductance of some monoatomic metallic wires have been observed by many authors [24, 25]. The conductance of a monoatomic chain of Au wire suspended between two gold tips found to be $G_0$ ($G_0 = 2e^2/h$: a quantum unit of conductance) [26]. Further investigations, however, have shown the conductance oscillations in Au wires [27, 28].

Oscillations in conductance of 1D linear Au wire with respect to the parity of the number of atoms N has been observed theoretically [29]. However, it has been shown that the equilibrium configurations of the wires are zigzag instead of linear structures, with essentially different electronic properties than those of linear ones [30-32]. The I-V curves of the atomic wires have been found to be strongly nonlinear. Conductance of gold wires on the basis of electronic structure calculations has been observed and addressed in several works [33-35]. These studies show that the chains have a tendency to dimerize upon strong elongation in accord with the Peierls mechanism. However, the conductance as a function of the chain length has not been studied consistently.

In this paper, transport properties of gold atomic wires are studied using density functional theory (DFT). Conductance, conductance pathways, transmission spectrum and IV characteristics atomic wires with different physical configurations, electrodes and length. In the past majority of the work is carried out using jallium electrodes []. In this work, two type of electrodes i.e. wire electrodes and semi-infinite crystalline electrodes are used to compare the former (ideal and most simplistic model) with the latter more realistic model.

## Computational Methods

First principle calculations were performed for gold monoatomic wires with varying number of atoms between three wires having different geometry and different kind of electrodes. Figure 1(a) shows gold atomic wire between wire-like electrodes. Figure 1 b and c show the straight and zigzag atomic wires embedded between semi-infinite crystalline electrodes respectively.

Density functional calculations were performed using local density approximation (LDA) with single zeta polarized basis set. Transmission is studied across the *c* axis shown in the figures.

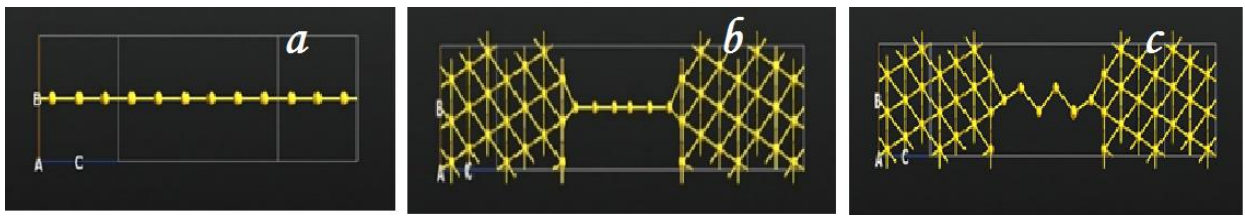

Figure 1. **(a)** Six atom straight wire embedded between the wire electrodes with 2.39 A interatomic distance. **(b)** Straight wire with 6 atoms embedded between semi-infinite gold electrodes with 2.39 A bond length. **(c)** Six atomic zigzag wire embedded between the semi-infinite gold electrodes. For both (b) and (c) electrodes are periodic in 'A' and 'B' direction whereas the transport is along the 'C' direction.

## Results and Discussion

Simulations were run to study the transmission properties of gold single atom wires connected between three different configurations as described in the introduction. The number of atoms in the wire ranges from three to eight with the interatomic spacing of 2.39Å to 2.69Å.

## Results of Stretching

We calculated the conductance for the wires in Figure 1a and Figure 1b by stretching the wires. Wires were stretched by increasing the bond length from 2.39Å to 2.69Å. It can be clearly seen that the conductance of the wire hanged between crystalline semiconductor electrodes monotonically approaches quantum conductance as the wire is stretched. On the other hand the conductance of the an atomic wire attached to wire-like electrodes show much higher conductance with irregular changes in the conductance upon stretching. Smooth interface at boundaries for the wire-like electrodes could be the reason for higher conductance and is further explored in the following text.

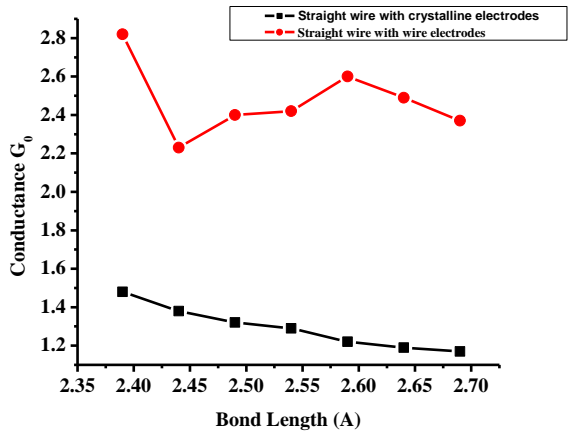

Figure 3. With the increase in bond length conductance is getting closer to the Quantum conductance ($G_0$) for the wire with semi-infinite gold electrodes. But for the long chain of atoms embedded between the wire electrodes, it shows abrupt changes.

## Conductance

Conductance were calculated or varying number of atomic lengths for straight wires with wire-like electrodes and semi-infinite crystalline electrodes and for the zig zag wire with semi-infinite crystalline electrodes. Laundaer-Buttiker equation was used to calculate the conductance which given below:

$$G = \frac{2e^2}{h} \sum_{i,j} T_{ij}(E_F) \qquad (1)$$

Transmission is calculated using retarded and advanced green functions $G^R(E)$ and $G^A(E)$ of the central scattering part. $T_{ij}(E_f, V_b)$= Tr[$\Gamma_L(E)$ $G^R(E)$ $\Gamma_R(E)$ $G^A(E)$ ]. Contact interface broadening at right and left contacts was included using terms $\Gamma_R(E)$ and $\Gamma_L(E)$.

For a zigzag wire with semi-infinite crystalline electrodes, even-odd oscillation is observed (D). Conductance of even number of atoms is below $1G_0$ while for odd number of atoms it is close to Quantum conductance $1G_0$. This behavior is similar to the one predicted by Smit. et al for chains of Na atoms between bulk electrodes using Friedel sum rule [41]. Tsukamoto and Hirose also verified this [42]. The linear chain of atoms suspended between semi-infinite crystalline electrodes also showed even-odd behavior (B). But oscillations are large, above Quantum conductance ($0.7G_0$-$1.5G_0$).

The linear chain of atoms with wire electrodes, conductance is linear at $2.32G_0$ for first three atoms, and then it increases. And again shows a linear behavior at $2.82G_0$. For further increase in number of atoms, conductance remained constant. Bond length for all three devices is same i.e. 2.39Å.

In the past, it has been reported that the conductance of chains of Au, Pt and Ir atoms oscillates as a function of number of atoms i.e. even-odd oscillations are observed [36]. Smit et al. [36] have discovered small variations of conductance from the value of $G_0$ in Au wires, whereas another study has reported extraordinarily large oscillations [37]. Conductance of gold wires on the basis of electronic structure calculations has been observed and addressed in several works [38-40]. These works show that the chains have a tendency to dimerize upon strong elongation in accord with the Peierls mechanism. However, the conductance as a function of the chain length has not been studied consistently.

Conductance of monoatomic wires suspended between two gold surfaces has been measured and is found to be very close to the Quantum conductance ($G_0$)[1].

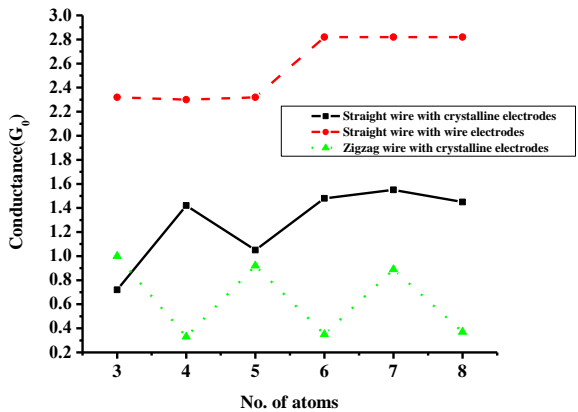

Figure 2. Conductance of monoatomic wires suspended between gold surfaces has been measured. $G_1$ shows the conductance of wire suspended between two semi-infinite electrodes. This wire clearly shows even-odd oscillations. $G_2$ shows the conductance of wire suspended between semi-infinite gold electrodes. $G_3$

## Transmission Spectrum

To get a better insight o the three different conductance behaviors transmission spectrums were calculated. Transmission spectrum was determined for an energy range of -6eV to 6eV. Figure 4 shows transmission spectrum for all three type of wires each consisting of six atoms. Figure 4(a) represent the simplest model. The structure is same in the electrode and in the scattering region hence a very smooth interface. Electron transmission above the Fermi-level has a transmission value of exactly 1 whereas at Fermi level (with a total transmission value 3) and below Fermi level even higher transmission is observed. Further analysis shows that there is exactly one transmission channel available above Fermi level whereas the multiple channels with significant transmission coefficient are available at and below Fermi level.

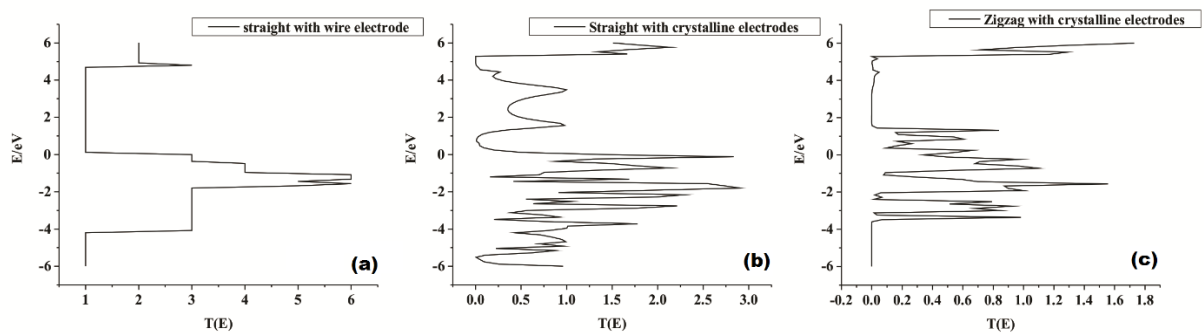

Figure 4 (a) straight atomic wire with wire like electrodes showing ideal behavior with quantized conductance. Conductance is observed to be 1, 3, 6 $G_0$ with three major contributing transmissions eigen-channels observed at Fermi level. (b) Transmission spectrum for straight wire with semi-infinite crystalline electrodes showing non-quantized transmission (c) Transmission Spectrum for zigzag wire with six number of atoms suspended between two semi-infinite crystalline electrodes.

Transmission spectrum of wire with semi-infinite gold electrodes and wire with atomic wire electrodes shows comparatively different behavior.

For both (straight and zig-zag) wires transmission is fairly suppressed. Analysis of individual eigen channel contribution shows that straight wire with crystalline show three major contributing channels at Fermi level. The transmission contributed by each channel is however suppressed (0.75, 0.34 and 0.34) is due to reflections in scattering region as well as the electrode-wire interface. In contrast to the straight wires there is only one transmission channel, for zig-zag wires consisting of both odd and even number of atoms. However, at Fermi level transmission coefficient of the single available eigen channels for even number atoms is much lower than odd number atom wires (0.34 compared to 0.97).

In order to explain why the transmission above Fermi level reduces abruptly in Figure 4(b) and Figure 4(c). Figure 5(a) and (b) shows calculated transmission pathways for the wire with wire-like-electrodes. It can been

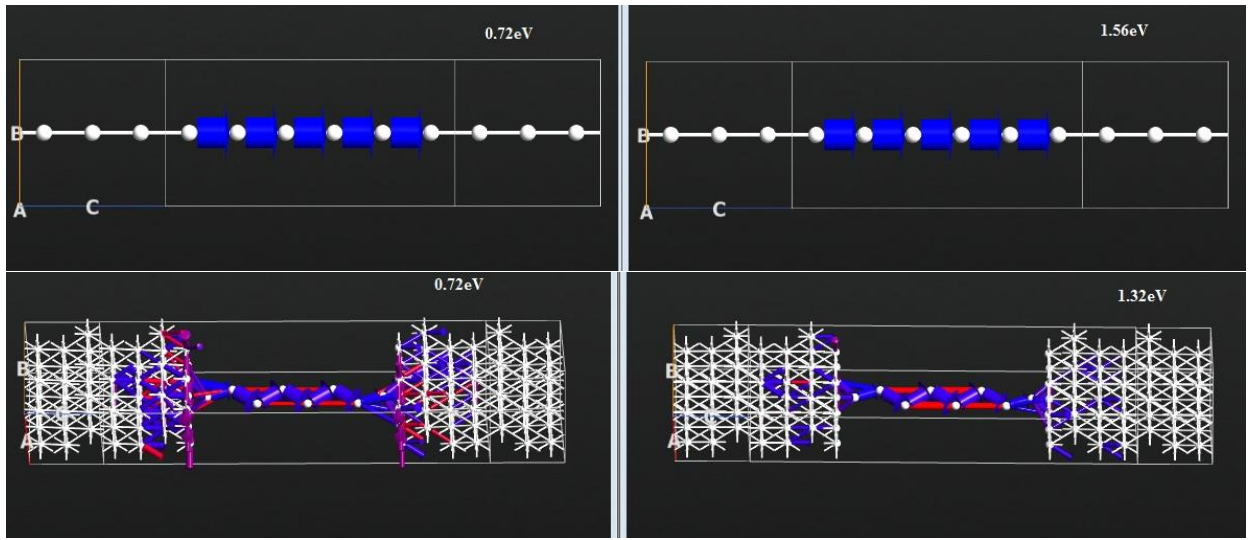

## Current-Voltage Curve

IV response was calculated using self-consistent technique. Transmission curves were calculated for each bias voltage and integrated self-consistently between the energies $-v/2$ and $v/2$. Figure 8 main panel shows the IV response for wires embedded in crystalline electrodes. The slope of the curves show clear regions with smaller slope and bigger slopes at regular intervals which is attributed to a periodic variation in conductance. However for wire embedded in wire-like electrodes show perfect ohmic behavior (Figure 8 inset).

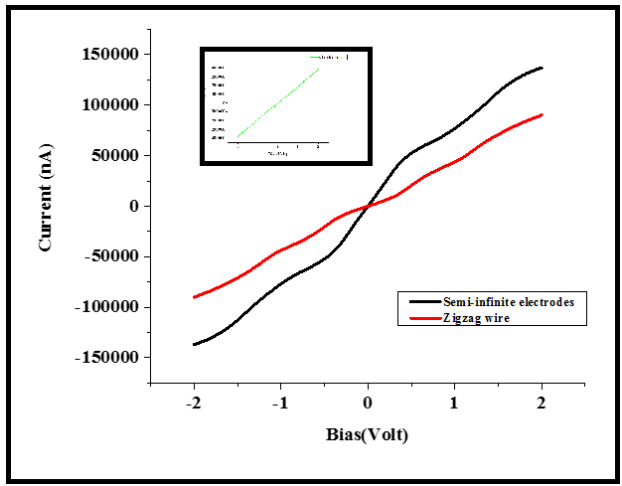

Figure 8: Computed Current-Voltage curves for three wires. Each having six numbers of atoms; one with semi-infinite electrodes (black), other with wire electrodes (red) and third with zigzag geometry (green). (Au-Au=2.39Å). Wire with zigzag geometry showed ohmic behavior, while there was small plateaus found in the IV's of straight and zigzag wire with semi-infinite bulk electrodes.

## Conclusion:

Electrical transport properties of straight and zig-zag gold atomic wires are explored in this work. A more realistic model with crystalline gold electrodes was simulated using DFT. Local density approximation (LDA) along with single zeta polarized (SZP) was used in the calculations. Simulations for wire-like ideal system were also run for comparison. It is clear from the results that the conductance of the model with crystalline electrodes is far less than the ideal case. The decrease in conductance is due to the suppression of transmission at many energies above Fermi level. This suppression is attributed to the reflection found at the interface of the wire and the electrodes. These reflections are nonexistent in the case where wire-like electrodes were used and the symmetry of the structure along the transmission axis is maintained throughout the device i.e. scattering region and electrodes.

## References:

[1]   U. Landman et al. 1990, Science 248, 454
[2]   Y. Kondo and K. Takayanagi 2000, Science 289, 606
[3]   Shinya Okano, Kenji Shiraishi, and Atsushi Oshiyama 2004, Physical Review B.
[4]   N. Agrait, A. Levy Yeyati, Jan M. Van Ruitenbeek 2003, Physics Reports, 81 – 279 377

[5]     Y. J. Lee, M. Brandbyge, M. J. Puska, J. Taylor, K. Stokbro, and R. M. Nieminen 2004, Phys. Rev. B 69, 125409
[6]     N. V. Skorodumova and S. I. Simak, Phys. Rev. B 67, 121404 (R) (2003)
[7]     N. V. Skorodumova, S. I. Simak, A. E. Kochetov, and B. Johansson, Phys. Rev. B 72, 193413 (2005)
[8]     N. Agrait, A. Yeyati, and J. van Ruitenbeek, Phys. Rep. 377, 81 (2003)
[9]     R. Smit, C. Untiendt, G. Rubio-Bollinger, R. Segers, and J. van Ruitenbeek, Phys. Rev. Lett. 91, 076805 (2003),
[10]   T. Kizuka, S. Umehara, and S. Fujisawa, Jpn. J. Appl. Phys. 40, L71 (2001)].
[11]   V. Rodrigues, T. Fuhrer, and D. Ugarte, Phys. Rev. Lett. 85, 4124 (2000)
[12]   C. Xiang, A. G. Guell, M. A. Brown, J. Y. Kim, J. C. Hemminger, and R. M. Penner, Nano Letters 8, 3017 (2008),
[13]   D. Azulai, T. Belenkova, H. Gilon, Z. Barkay, and G. Markovich, Nano Letters 9, 4246 (2009)]
[14]   Q.-Q. Wang, J.-B. Han, D.-L. Guo, S. Xiao, Y.-B. Han, H.-M. Gong, and X.-W. Zou, Nano Letters 7, 723 (2007)]
[15]   X. Zhang, H. Liu, J. Tian, Y. Song, and L. Wang, Nano Letters 8, 2653 (2008)
[16]   F. Patolsky and C. M. Lieber, Materials Today 8, 20 (2005)
[17]   U. Yogeswaran and S.-M. Chen, Sensors 8, 290 (2008)
[18]   Z. Liu and P. C. Searson, The Journal of Physical Chemistry B 110, 4318 (2006)
[19]   X. Wang and C. S. Ozkan, Nano Letters 8, 398 (2008)
[20]   S. Huang and Y. Chen, Nano Letters 8, 2829 (2008).
[21]   M. R. Sorensen, M. Brandbyge, and K. W. Jacobsen, Phys. Rev. B 57, 3283 (1998)
[22]   H. Ohnishi, Y. Kondo, and K. Takayanagi, Nature (London) 395, 780 (1998)
[23]   A. I. Yanson et al., Nature (London) 395, 783 (1998)].
[24]   R. H. M. Smit, C. Untied, G. Rubio- Bollinger, R. C. Segers, and J. M. Ruitenbeek, Phys. Rev. Lett. 91, 076805 (2003)
[25]   Yan-hong Zhou, Xiao-hong Zheng, Ying Xu and Zhao Yang Zeng, J. Phys.: Condens. Matter 20 (2008), 045225].
[26]   N. Agrait, A. Yeyati, and J. van Ruitenbeek, Phys. Rep. 377, 81 (2003)
[27] J. Costa-Kramer, N. Garcia, P. Garcia- Movhales, P. Serena, M. Marques, and A. Correia, Phys. Rev. B 55, 5416 (1997)
[28]   T. Kizuka, S. Umehara, and S. Fujisawa, Jpn. J. Appl. Phys. 40, L71 (2001)
[29]   Y. J. Lee, M. Brandbyge, M. J. Puska, J. Taylor, K. Stokbro, and R. M. Nieminen, Phys. Rev. B. 69, 125409 (2004)
[30]   N. Agrait, A. Yeyati, and J. van Ruitenbeek, Phys. Rep. 377, 81 (2003)
[31]   Y. J. Lee, M. Brandbyge, M. J. Puska, J. Taylor, K. Stokbro, and R. M. Nieminen, Phys. Rev. B. 69, 125409 (2004)
[32] N. V. Skorodumova, S. I. Simak, A. E. Kochetov, and B. Johansson, Phys. Rev. B 72, 193413 (2005).
[33]   M. Okamoto and K. Takayanagi, Phys. Rev. B 60, 7808 (1999)
[34]   H. Hakkinen, R.N. Barnett, and U. Landman, J. Phys. Chem. B 103, 8814 (1999)
[35]   T. Ono, H. Yamasaki, Y. Egami, and K. Hirose, Cond- mat/ 0212603 (unpublished)

[36] R.H.M. Smit, C. Untiedt, G. Rubio-Bollinger, R.C. Segers, and J.M. Ruitenbeek, Phys. Rev. Lett.91, 076805 (2003)

[37]. J. Costa-Kara ̈mer, N. Garcia, P. Garcia-Movhales, P. Serena, M. Marques, and A. Correia, Phys. Rev. B55, 5416 (1997).

[38]. M. Okamoto and K. Takayanagi, Phys. Rev. B60, 7808 (1999)

[39] H. Hakkinen, R.N. Barnett, and U. Landman, J. Phys. Chem. B 103, 8814 (1999)

[40] T. Ono, H. Yamasaki, Y. Egami, and K. Hirose, Cond-mat/0212603(unpublished)[51] H.S. Sim, H.W. Lee, and K.J. Chang, Phys. Rev. Lett.87, 096803(2001)

[41] M. Brandbyge, N. Kobayashi, and M. Tsukada, Phys. Rev. B60, 17 064(1999)

[42] S. Tsukamoto and K. Hirose, Phys. Rev. B66, 161402 (2002)